# High-pressure growth and characterization of bulk MnAs single crystals

Nikolai D. Zhigadlo

*Department of Chemistry and Biochemistry, University of Bern, Freiestrasse 3, 3012 Bern, Switzerland*

**Abstract**

Bulk single crystals of manganese arsenide (MnAs) were grown from melt at 1 GPa and 1100 °C by using a cubic-anvil, high-pressure, and high-temperature technique. The as-grown black colored crystals extracted from solidified lump exhibit a plate-like morphology, with flat surfaces and maximum dimensions up to ~ 3 × 2 × 0.5 mm$^3$. The hexagonal crystal structure at room temperature was confirmed by X-ray diffraction [B8$_1$, space group $P6_3/mmc$, No 194, $Z$ = 2, $a$ = 3.7173(4) Å, $b$ = 3.7173(4) Å, $c$ = 5.7054(8) Å, and V = 68.277(16) Å$^3$]. Temperature-dependent magnetization measurements reveal the occurrence of a first-order ferro- to paramagnetic transition at $T_c$ = 318.5 K accompanied by a hysteresis of ~ 9 K. The successful growth of relatively large crystals reported here might be extended to various substituted analogues of MnAs, thus opening new possibilities for further exploration of this interesting system.

Corresponding author.

*E-mail address*: nikolai.zhigadlo@dcb.unibe.ch, nzhigadlo@gmail.com (N. D. Zhigadlo)

## 1. Introduction

The family of transition-metal compounds with a general formula MX (M = transition metal, X = P, As, Sb) represent a large class of materials with a wealth of stoichiometries, structures, and physical properties including semiconductivity, ferromagnetism, thermoelectricity, and superconductivity [1-5]. Among them, manganese arsenide (MnAs), which hosts a variety of magnetic orderings, has been extensively studied for many decades [6,7]. At ambient pressure, bulk MnAs shows two successive phase transitions as the temperature is increased (Fig. 1) [8]. A first-order phase transition from the hexagonal ferromagnetic (FM) α-MnAs (NiAs-type, $P6_3/mmc$, (see Fig. 2)) to the paramagnetic (PM) β-MnAs (MnP-type, $Pnma$) phase occurs at $T_c \sim 317$ K [9]. Because of the *magnetostructural* coupling, this first-order magnetic phase transition (α→β) is accompanied by a discontinuous change in volume (~ 2%), resistivity, heat capacity and magnetization [6,10,11]. At an even higher temperature, ~ 393 K, the crystal structure returns to the hexagonal NiAs-type structure (γ-MnAs) with $P6_3/mmc$ space group, through the second-order phase transition, maintaining, however, the paramagnetic state.

In recent years MnAs has attracted renewed attention due to a giant magnetocaloric effect (MCE) [12,13], as well as for applications related to information and energy storage [14]. Notably, MnAs shows comparable MCE properties to those of the rare-earth compound $Gd_2(Si_2Ge_2)$, where a giant MCE was first reported by Pecharsky and Gschneidner [15]. The maximum entropy change in MnAs can reach $\Delta S = S_\alpha - S_\beta = -30$ J (K$^{-1}$ kg$^{-1}$) for a magnetic field change $\Delta H = 5$ T and -20 J (K$^{-1}$ kg$^{-1}$) for $\Delta H = 2$ T [12]. These huge entropy variations reflect its complex magnetostructural phase transitions.

Interestingly, the unusual properties of bulk MnAs are preserved in MnAs films too, thus extending the possible applications of the material in spintronic devices (see ref. [16] for overview). MnAs has attracted great attention also in combination with GaAs, either as a MnAs/GaAs hybrid system or as a ternary dilute semiconductor (Ga,Mn)As. The Curie temperature of such GaAs:Mn/MnAs hybrid systems can reach values of up to 340 K. The above mentioned MnAs properties, together with the possibilities offered by hybrid systems open up new venues towards the realization of novel devices based on a planar geometry [14,16].

To understand and explain the origin of the many interesting properties of MnAs, and to estimate its potential for practical applications, it is preferable starting with studies of single crystal samples. However, from a critical review of the published results one promptly realizes that MnAs is not easy to grow. Various obstacles are encountered in the synthesis of MnAs, the main issue being the high vapor pressure of As, which requires stringent safety precautions. In this paper we report an alternative way, based on a high-pressure and high-temperature (HPHT) method for growing high-quality bulk single crystals of relatively large sizes. Besides the details of crystal growth, we report on their structural and magnetic characterization.

Before going into the details of HPHT growth of MnAs, let us first recall some key points regarding earlier growth attempts. Several techniques have been implemented to prepare crystalline MnAs, namely: a) by synthesis of the elements under high arsenic pressure, while keeping the temperature far below the MnAs melting point; b) through alloying the elements by heating them under varying arsenic pressures up to temperatures above the melting point of MnAs [6, 7]. However, both these methods yield crystals of only a few tenths of a millimeter. Paitz [17] implemented a horizontal zone-melting technique for the growth of MnAs crystals and discussed the difficulties arising from the polymorphic transformations of the material, leading to the fracturing of MnAs into pencil-like grains. Bärner and Berg [18] conducted the synthesis around 935 °C by filling the mixed powders into a quartz ampoule, which was then inserted into a steel case whose free volume was filled with quartz powder. The main product was polycrystalline MnAs accompanied by rod-shaped whiskers. Then Govor et al. [19] described the preparation of MnAs crystals using the Bridgman technique and obtained crystals with dimensions up to $0.1 \times 0.3 \times 2$ mm$^3$. Later on, De Campos et al. [20] failed to reproduce the growth via Bridgman method, but succeeded via application of a complicated heat-treatment protocol in a vertical furnace. In summarizing all the previous attempts we note that a direct synthesis of MnAs requires much caution to avoid a high arsenic pressure. The vapour pressure of pure arsenic rises very rapidly as temperature increases, reaching 37 atm at 818 °C [17]. Consequently, the temperature has to be increased very slowly for the reaction to be completed before the arsenic vapor pressure becomes too high. Also, it has been noticed that both, the quartz and alumina crucibles show a tendency to be wet by MnAs. Cracking of the wall of quartz tube due to reaction of manganese with quartz represents also a serious obstacle. The high-pressure synthesis reported below allows us to completely eliminate all these drawbacks.

## 2. Experimental details

For the growth of MnAs single crystals, we used the cubic-anvil, high-pressure, and high-temperature technique. The details of experimental set up can be found in our previous publications [5,21]. The apparatus consist of a 1500-ton press, with a hydraulic-oil system comprising a semi-cylindrical multianvil module (Rockland Research Corp.). A set of steel parts transmit the force through six tungsten carbide pistons to the sample in a quasi isostatic way. This method was successfully applied earlier on to grow various compounds, including superconducting intermetallic crystals [22-24], diamonds [25], cuprate oxides [26,27], pyrochlores [28], *Ln*Fe*Pn*O (*Ln*1111, *Ln*: lanthanide, *Pn*: pnictogen) oxypnictides [29-31], and numerous other compounds [32,33]. The limited availability of high-pressure phase diagrams is fully compensated by a number of advantages offered by the use of high-pressure conditions.

A mixture of stoichiometrically equal amounts of manganese powder (purity 99.99 %) and arsenic powder (purity 99.9999 %) was thoroughly ground and pressed into a pellet of 8 mm in diameter and 8 mm in length. The pellet was then placed in a boron-nitride (BN) container surrounded by a graphite-sleeve resistance heater and inserted into a pyrophyllite cube. The temperature was calibrated in advance by using a B-type PtRh6% thermocouple and related to the power dissipated in the pressure cell. To avoid overheating the tungsten carbide anvils, a water cooling system was installed. In order to minimize a possible degradation or contamination of the material due to air exposure, all the work related to the sample preparation and the packing of the high-pressure cell-assembly was performed in a glove box with a protective argon atmosphere. In the high-pressure synthesis of MnAs the assembled cell was compressed to 1 GPa at room temperature, and the optimum growth conditions were tuned by varying the heating temperature, the reaction time, and the cooling rate. After this preliminary optimization, we used a synthesis temperature of about 1100 °C, which was found to be optimal for growing sizable MnAs single crystals. In a typical growth process the BN crucible was heated up to ~ 1100 °C in 3 h and maintained there for 2 h to assure melting and a thorough mixing of the components. After that the melt was slowly cooled to 900 °C over 10 h and finally the crucible was quenched to room temperature before releasing pressure. These environmental conditions and synthesis protocol avoid the fracturing of MnAs crystals into prismatic grains, which is often observed at ambient-pressure synthesis due to low temperature polymorphic transformations. The final product inside the crucible was almost fully melted into a cylindrically-shaped solidified lump. After completing the crystal growth process, the MnAs crystals were mechanically extracted from the solidified lump.

The X-ray single crystal diffraction measurements were performed at room temperature on an *Oxford Diffraction SuperNova* area-detector diffractometer [34] using mirror optics, monochromatic Mo $K_\alpha$ radiation ($\lambda = 0.71073$ Å), and Al filter [35]. The unit cell parameters and the orientation matrix for the data collection were obtained from a least-squares refinement, using reflections angles in the range $3.5 < \theta < 31.3°$. A total of 728 frames were collected using ω scans, with 5+5 seconds exposure time, a rotation angle of 1.0° per frame, and a crystal-detector distance of 65.0 mm at $T = 298(2)$ K. The data reduction was performed using the *CrysAlisPro* [34] program. The intensities were corrected for Lorentz and polarization effects, and an absorption correction based on the multi-scan method using SCALE3 ABSPACK in *CrysAlisPro* [34] was applied. Large-size crystals were measured in Bragg-Brentano geometry on a STOE diffractometer using monochromatic Cu $K_\alpha$ radiation ($\lambda = 1.54060$ Å). The elemental analysis of the grown crystals was performed by means of Energy Dispersive X-ray spectroscopy (EDX, Hitachi S-3000 N). The temperature-dependent magnetization measurements were carried out using a Magnetic Property Measurement System (MPMS-XL, Quantum Design) equipped with a reciprocating-sample option.

3. **Results and discussion**

When heated, arsenic does not melt under ambient pressure, as most solids do, but sublimes at 614 °C directly into gas [36]. However, under high pressure, it melts and the solid-liquid equilibria were determined up to 7 GPa [36]. Extrapolating of these data to 1.013 bar yields 817 °C as the melting point of the hypothetically existing solid. With increasing pressure the melting point of As increases with a rate of ~4 °C/kbar reaching a value of ~855 °C at 10 kbar (1 GPa). For the Mn-As system at the temperatures higher than 855 °C the liquid arsenic most probably works as a flux promoting the solution of the Mn constituent. Thus, the combined 1 GPa – 1100 °C conditions used in this work seem very favorable for growing bulk MnAs single crystals and our results suggest that MnAs melts congruently not only at ambient-pressure (~935 °C) but also at high-pressure conditions. To further clarify the mechanism of MnAs growth one needs to know the detailed pressure-temperature phase diagram of the Mn-As system, which currently is not available.

The extracted black colored MnAs crystals were found to exhibit irregular plate-like shapes with flat surfaces reaching maximum dimensions of ∼ 3 × 2 × 0.5 mm$^3$ (Fig. 3 inset). The EDX analysis performed on several relatively large crystals shows a homogeneous distribution of Mn and As atoms. The ratio of Mn-to-As atoms was determined by averaging 3-5 measurements and found to be 1:1, consistent with the stoichiometric composition, within an experimental error of ∼ 0.01. A large-size MnAs crystal was aligned to the surface of a sample holder and attached there by using vacuum grease. Fig. 3 shows an X-ray diffraction pattern taken from the flat surface confirming the single crystalline nature of the sample. The presence of diffraction lines with (*h00*) indices indicates that the flat surface corresponds to the *bc*-plane, with *a*-axis being perpendicular to the face.

Let us now focus on the structural details of the grown crystals. The first single crystal MnAs structure was reported in the 60's [9]. To check the structure of our crystals we collected diffraction data on several MnAs pieces, originating from different growth batches. For each case, a useful single crystal could be found and the refined structural model was essentially equivalent to the reported one [9].

A full X-ray refinement was performed at room temperature on a crystal with dimensions 0.35 × 0.15 × 0.06 mm$^3$, using 643 reflections (of which, 59 unique) in the *k*-space region $-5 \leq h \leq 5$, $-5 \leq k \leq 5$, $-8 \leq l \leq 8$, by a full-matrix least-squares minimization of $F^2$. The weighting scheme was based on counting statistics and included a factor to down-weight the most intense reflections. The structure refinement was performed using the SHELXL-2014 program [37], in the hexagonal space *P*6$_3$/*mmc*. Fig. 2 shows a 3D view of the crystal structure of MnAs with unit cell dimensions $a = b = 3.7173(4)$ Å and $c = 5.7054(8)$ Å at room temperature. The hexagonal NiAs-type structure is one of the most common for MX-type structures. The magnetic Mn atoms in the Wyckoff position 2*a* (0 0 0) occur in plane sheets, normal to the *c*-axis and with *c*/2 spacing, while As atoms are located in the 2*c* (1/3 2/3 1/4) positions, thus forming a hexagonal close-packed array. The crystallographic parameters are summarized in Table 1. The refined atomic positions, bond lengths, and angles are presented in Table 2 and 3. The full thermal displacement parameters $U^{ij}$ for all the atoms are reproduced in Table 4. The $R_1$ refinement factor was 2.93% with w$R_2$ = 7.19% including all data. The chemical formula of MnAs was confirmed from the fully occupied positions for all sites. Further evidence on the 1:1 stoichiometry is obtained from the *c/a* ratio. It is known, that high *c/a* values (~ 1.4 – 1.6) correspond to compounds which exhibit a 1:1 stoichiometry, whereas lower *c/a* values (~ 1.2 – 1.3) correspond to Mn-rich compounds. Since the *c/a* ration of our MnAs crystals is 1.53, it confirms the stoichiometric composition of the grown crystals. Fig. 4 shows the image of a single-crystal MnAs with unit cell axes and the reconstructed *0kl*, *h0l*, and *hk0* reciprocal-space sections measured at room temperature. Well-resolved reflections confirm the high quality of the single crystal used for the structural study. No additional phases, impurities, twins, or intergrowing crystals were detected.

Figure 5 shows the temperature-dependent magnetic response of a MnAs single crystal in a field of 10 Oe with increasing (ZFC) and decreasing temperatures (FC). The hysteresis in the temperature-driven first-order transition ($T_{hys}$ = 9 K) with characteristic transition temperatures ($T_c^{FM-PM}$ = 318.5 K and $T_c^{PM-FM}$ = 309.5 K) is clearly observed. At low temperatures the divergence between the FC and ZFC curves is visible. This type of discrepancy is very common in spin glass, cluster spin glass, and superparamagnets. However, for the present ferromagnetic system where long-range magnetic ordering is involved, the upturn in the FC curve can arise from domain-wall motions and/or disorder [38,39]. Cooling under the applied magnetic field favors domain growth in the direction of applied field, thus giving higher value of magnetization. On the other hand, domain growth will be more random in the case of zero-field cooling and will be governed by magnetocrystalline anisotropy, giving low net magnetization. One of the additional contributing factors towards the observed upturn in the FC branch could be due to the inherent structural disorder.

While the sharp transition temperatures can be considered as a sign of high-quality material, a large $T_{hys}$ is not promising for refrigeration applications. In fact, for practical applications one should be able to retain the first-order transition, while at the same time having a low hysteresis. Various microstructural features, such as the crystal size and shape, porosity, secondary phases, inhomogeneity, and strain can change significantly the path along which this structural change actually occurs. A subtle derivation in the transition mechanism suffices to bring about a dramatic variations in the critical temperature of the coupled first-order phase transitions and consequently it can substantially affect the hysteresis. A weakening of the first-order transition by compositional modification (i.e., via substitutions of Mn by Cr or Cu, and As by P) is a common method to reduce the hysteresis, however, as a side effect, the resultant entropy change can also be significantly lowered [40, 41]. The reduction of the detrimental hysteresis requires, therefore, a clear strategy regarding an appropriate microstructure design, in turn requiring a better understanding of the magnetic phase transitions in MnAs.

## 4. Conclusion

A new method to grow single crystals of MnAs was presented. By using a cubic-anvil, high-pressure apparatus, bulk single crystals of MnAs were grown from stoichiometric melt at 1100 °C under a pressure of 1 GPa. The crystal structure of MnAs was confirmed by single-crystal X-ray diffraction. Electron-probe microanalysis (EDX) indicates a homogeneous distribution of Mn and As atoms across the crystals. Temperature-dependent magnetization measurements reveal the occurrence of a first-order ferro- to paramagnetic phase transition at 318 K with a pronounced thermal hysteresis (~ 10 K). The successful growth of MnAs crystals demonstrated here could also be extended to the synthesis of various substituted analogues, thus opening up new possibilities for the further exploration of these interesting materials.

**Acknowledgements**

I would like to acknowledge P. Macchi for his help in performing the structural characterization and J. Hulliger and T. Shiroka for critically reading the manuscript and for helpful comments. The Swiss National Science Foundation is acknowledged for co-funding the single-crystal X-ray diffractometer (R-equip project 206021_128724).

# References

[1] A. Continenza, S. Picozzi, W.T. Geng, A.J. Freeman, Phys. Rev. B 64 (2001) 085204.

[2] B. Saparov, J.E. Mitchell, A.S. Sefat, Supercond. Sci. Technol. 25 (2012) 084016.

[3] a) R. Khasanov, A. Amato, P. Bonfa, Z. Guguchia, H. Luetkens, E. Morenzoni, R. De Renzi, N.D. Zhigadlo, Phys. Rev. B 93 (2016) 180509 (R); b) R. Khasanov, A. Amato, P. Bonfa, Z. Guguchia, H. Luetkens, E. Morenzoni, R. De Renzi, N.D. Zhigadlo, J. Phys.: Condens Matter 29 (2017) 164003.

[4] R. Khasanov, Z. Guguchia, I. Eremin, H. Luetkens, A. Amato, P. K. Biswas, Ch. Rüegg, M.A. Susner, A.S. Sefat, N.D. Zhigadlo, E. Morenzoni, Sci. Rep. 5 (2015) 13788.

[5] N.D. Zhigadlo, N. Barbero, T. Shiroka, J. Alloy Compd. 725 (2017) 1027-1034.

[6] B.T.M. Willis, H.P. Rooksby, Proc. Phys. Soc. London B 67 (1954) 290.

[7] R.W. De Blois, D.S. Rodbell, Phys. Rev. 130 (1963) 1347.

[8] H. Okamoto, Bull. Alloy Phase Diagrams, 10 (1989) 549-554.

[9] R.H. Wilson, J.S. Kasper, Acta Cryst. 17 (1964) 95.

[10] F. Grønvold, S. Snidal, Acta Chem. Scand. 24 (1970) 285-298.

[11] A. Zięba, K. Selte, A. Kjekshus, A.F. Andresen, Acta Chem. Scand. A32 (1978) 173-177.

[12] H. Wada, Y. Tanabe, Appl. Phys. Lett. 79 (2001) 3302.

[13] S. Gama, A.A. Coelho, A. de Campos, A. Magnus, G. Carvalho, F.C.G. Gandra, P.J. von Ranke, N.A. de Oliveira, Phys. Rev. Lett. 93 (2004) 237202.

[14] J. Lyubina, J. Phys. D: Appl. Phys. 50 (2017) 053002.

[15] V.K. Pecharsky, K.A. Gschneidner, Jr., Phys. Rev. Lett. 78 (1997) 4494.

[16] L. Däweritz, Rep. Prog. Phys. 69 (2006) 2581-2629.

[17] J. Paitz, J. Cryst. Growth 11 (1971) 218-220.

[18] K. Bärner, H. Berg, J. Cryst. Growth 46 (1979) 763-770.

[19] a) G.A. Govor, Phys. Stat. Sol (a) 91 (1986) K59; b) V.I. Mitsiuk, N.Yu. Pankratov, G.A. Govor, S.A. Nikitin, A.I. Smarzhevskaya, Phys. Solid State 54 (2012) 1988-1995.

[20] A. De Campos, M.A. Mota, S. Gama, A.A. Coelho, B.D. White, M.S. da Luz, J.J. Neumeier, J. Cryst. Growth 333 (2011) 54-58.

[21] a) N.D. Zhigadlo, S. Katrych, Z. Bukowski, S. Weyeneth, R. Puzniak, J. Karpinski, J. Phys.: Condens Matter 20 (2008) 342202; b) N.D. Zhigadlo, S. Weyeneth, S. Katrych, P.J.W. Moll, K. Rogacki, S. Bosma, R. Puzniak, J. Karpinski, B. Batlogg, Phys. Rev. B 86 (2012) 214509.

[22] a) R.T. Gordon, N.D. Zhigadlo, S. Weyeneth, S. Katrych, R. Prozorov, Phys. Rev. B 87 (2013) 094520; b) D. Ernsting, D. Billington, T. Millichamp, R. Edwards, H. Sparkes, N.D. Zhigadlo, S. Giblin, J. Taylor, J. Duffy, S. Dugdale, Sci. Reports 7 (2017) 10148.

[23] a) J. Karpinski, N.D. Zhigadlo, S. Katrych, R. Puzniak, K. Rogacki, R. Gonnelli, Physica C 456 (2007) 3-13; b) N.D. Zhigadlo, S. Katrych, J. Karpinski, B. Batlogg, F. Bernardini, S. Massida, R. Puzniak, Phys. Rev. B 81 (2010) 054520.

[24] N.D. Zhigadlo, J. Cryst. Growth 455 (2016) 94-98.

[25] N.D. Zhigadlo, J. Cryst. Growth 395 (2014) 1-4.

[26] a) N.D. Zhigadlo, J. Karpinski, Physica C 460 (2007) 372-373; b) R. Khasanov, N.D. Zhigadlo, J. Karpinski, H. Keller, Phys. Rev. B 79 (2007) 094505.

[27] a) N.D. Zhigadlo, Y. Anan, T. Asaka, Y. Ishida, Y. Matsui, E. Takayama-Muromachi, Chem. Mat. 11 (1999) 2185-2190; b) E. Takayama-Muromachi, T. Drezen, M. Isobe, N.D. Zhigadlo, K. Kimoto, Y. Matsui, E. Kita, J. Solid State Chem. 175 (2003) 366-371; c) N.D. Zhigadlo, K. Kimoto, M. Isobe, Y. Matsui, E. Takayama-Muromachi, J. Solid State Chem. 170 (2003) 24-29.

[28] a) S.M. Kazakov, N.D. Zhigadlo, M. Bruhwiler, B. Batlogg, J. Karpinski, Supercond. Sci. Technol. 17 (2004) 1169-1172; b) S. Katrych, Q.F. Gu, Z. Bukowski, N.D. Zhigadlo, G. Krauss, J. Karpinski, J. Solid State Chem. 182 (2009) 428-434.

[29] a) N.D. Zhigadlo, S. Katrych, S. Weyeneth, R. Puzniak, P.J.W. Moll, Z. Bukowski, J. Karpinski, H. Keller, B. Batlogg, Phys. Rev. B 82 (2010) 064517; b) N.D. Zhigadlo, S. Katrych, M. Bendele, P.J.W. Moll, M. Tortello, S. Weyeneth, V.Yu. Pomjakushin, J. Kanter, R. Puzniak, Z. Bukowski, H. Keller, R.S. Gonnelli, R. Khasanov, J. Karpinski, B. Batlogg, Phys. Rev. B 84 (2011) 134526.

[30] N.D. Zhigadlo, J. Cryst. Growth 382 (2013) 75-79.

[31] N.D. Zhigadlo, M. Iranmanesh, W. Assenmacher, W. Mader, J. Hulliger, J. Supercond. Nov. Magn. 30 (2017) 79-84.

[32] C. Müller, N.D. Zhigadlo, A. Kumar, M.A. Baklar, J. Karpinski, P. Smith, T. Kreouzis, N. Stingelin, Macromolecules 44 (2011) 1221-1225.

[33] N.D. Zhigadlo, J. Cryst. Growth 402 (2014) 308-311.

[34] Oxford Diffraction, CrysAlisPro, Oxford Diffraction Ltd. Yarnton, Oxfordshire, UK, 2010, Version 1.171.34.44.

[35] P. Macchi, H.-B. Bürgi, A.S. Chimpri, J. Hauser, Z. Gal, J. Appl. Cryst. 44 (2011) 763-771.

[36] N.A. Gokcen, Bull. Alloy Phase Diagrams 10 (1989) 11-22.

[37] a) G.M. Sheldrick, Acta Cryst. A71 (2015) 3-8; b) G.M. Sheldrick, Acta Cryst. C71 (2015) 3-8.

[38] S. M. Yusuf, L. Madhar Rao, J. Phys.: Condens. Matter 7 (1995) 5891.

[39] Amit Kumar, S. M. Yusuf, L. Keller, Phys. Rev. B 71 (2005) 054414.

[40] V.K. Pecharsky, K.A. Gschneidner Jr., Int. J. Refrig. 29 (2006) 1239-1249.

[41] E. Brück, O. Tegus, D.T. Cam Thanh, Nguyen T. Trung, K.H.J. Buschow, Int. J. Refrig. 31 (2008) 763-770.

**Table 1.** Details of single-crystal X-ray diffraction data and crystal refinement results for MnAs.

| | |
|---|---|
| Identification code | shelx |
| Empirical formula | MnAs |
| Formula weight | 129.86 g/mol |
| Temperature | 298(2) K |
| Wavelength | Mo K$_\alpha$ (0.71073 Å) |
| Crystal system | Hexagonal |
| Space group | *P6$_3$/mmc* |
| Unit cell dimensions | $a = 3.7173(4)$ Å, $\alpha = 90°$ |
| | $b = 3.7173(4)$ Å, $\beta = 90°$ |
| | $c = 5.7054(8)$ Å, $\gamma = 120°$ |
| Cell volume | 68.277(16) Å$^3$ |
| Z | 2 |
| Density (calculated) | 6.317 Mg/m$^3$ |
| Absorption coefficient | 32.925 mm$^{-1}$ |
| F(000) | 116 |
| Crystal size | 0.3532 × 0.146 × 0.0556 mm$^3$ |
| $\theta$ range for data collection | 6.338 - 31.397° |
| Index ranges | $-5 \leq h \leq 5, -5 \leq k \leq 5, -8 \leq l \leq 8$ |
| Reflections collected | 643 |
| Independent reflections | 59 [$R_{int} = 0.0867$] |
| Completeness to $\theta = 25.000°$, % | 100.0 % |
| Absorption correction | Gaussian |
| Max. and min. transmission | 0.177 and 0.032 |
| Refinement method | Full-matrix least-squares on $F^2$ |
| Data / restraints / parameters | 59 / 0 / 6 |
| Goodness-of-fit on $F^2$ | 1.213 |
| Final R indices [$I > 2\sigma(I)$] | $R_1 = 0.0276$, w$R_2 = 0.0698$ |
| R indices (all data) | $R_1 = 0.0293$, w$R_2 = 0.0719$ |
| Extinction coefficient | 0.14(3) |
| Largest diff. peak and hole | 1.002 and -0.682 e.Å$^{-3}$ |

**Table 2.** Position atomic coordinates ( $\times 10^4$ ), in space group $P6_3/mmc$ at ambient temperature and pressure, and their equivalent isotropic displacement parameters ($Å^2 \times 10^3$) for MnAs. $U$(eq) is defined as one third of the trace of the orthogonalized $U^{ij}$ tensor.

| Atom | Wyckoff | x | y | z | U(eq) |
|---|---|---|---|---|---|
| Mn1 | 2a | 0 | 0 | 0 | 15(1) |
| As(1) | 2c | 3333 | 6667 | 2500 | 15(1) |

**Table 3.** Bond lengths (Å) and angles (°) for MnAs.

___________________________________________________________________

| | | | |
|---|---|---|---|
| Mn1-As1#1 | 2.5769(2) | As1#2-Mn1-Mn1#6 | 56.392(5) |
| Mn1-As1#2 | 2.5769(2) | As1#3-Mn1-Mn1#6 | 123.608(5) |
| Mn1-As1#3 | 2.5769(2) | As1-Mn1- Mn1#6 | 123.608(4) |
| Mn1-As1 | 2.5769(2) | As1#4-Mn1-Mn1#6 | 56.392(4) |
| Mn1-As1#4 | 2.5770(2) | As1#5-Mn1-Mn1#6 | 123.608(4) |
| Mn1-As1#5 | 2.5770(2) | As1#1-Mn1-Mn1#7 | 123.608(4) |
| Mn1-Mn1#6 | 2.8527(4) | As1#2-Mn1-Mn1#7 | 123.608(5) |
| Mn1-Mn1#7 | 2.8527(4) | As1#3-Mn1-Mn1#6 | 56.392(5) |
| As1-Mn1#8 | 2.5769(2) | As1-Mn1-Mn1#7 | 56.392(4) |
| As1-Mn1#9 | 2.5769(2) | As1#4-Mn1-Mn1#7 | 123.608(4) |
| As1-Mn1#10 | 2.5769(2) | As1#5-Mn1-Mn1#7 | 56.392(4) |
| As1-Mn1#11 | 2.5769(2) | Mn1#6-Mn1-Mn1#7 | 180.0 |
| As1-Mn1#7 | 2.5769(2) | Mn1#8-As1-Mn1 | 130.782(3) |
| | | Mn1#8-As1-Mn1#9 | 130.782(3) |
| As1#1-Mn1-As1#2 | 92.319(6) | Mn1-As1-Mn1#9 | 92.318(7) |
| As1#1-Mn1-As1#3 | 87.681(6) | Mn1#8-As1-Mn1#10 | 92.318(7) |
| As1#2-Mn1-As1#3 | 180.0 | Mn1-As1-Mn1#10 | 130.782(2) |
| As1#1-Mn1-As1 | 180.0 | Mn1#9-As1-Mn1#10 | 67.216(8) |
| As1#2-Mn1-As1 | 87.682(6) | Mn1#8-As1-Mn1#11 | 67.216(9) |
| As1#3-Mn1-As1 | 92.318(6) | Mn1-As1-Mn1#11 | 92.318(6) |
| As1#1-Mn1-As1#4 | 92.318(6) | Mn1#9-As1-Mn1#11 | 92.318(6) |
| As1#2-Mn1-As1#4 | 92.318(6) | Mn1#10-As1-Mn1#11 | 130.782(2) |
| As1#3-Mn1-As1#4 | 87.682(6) | Mn1#8-As1-Mn1#7 | 92.318(6) |
| As1-Mn1-As1#4 | 87.682(6) | Mn1-As1-Mn1#7 | 67.216(9) |
| As1#1-Mn1-As1#5 | 87.682(6) | Mn1#9-As1-Mn1#7 | 130.782(2) |
| As1#2-Mn1-As1#5 | 87.682(6) | Mn1#10-As1-Mn1#7 | 92.318(6) |
| As1#3-Mn1-As1#5 | 92.318(6) | Mn1#11-As1-Mn1#7 | 130.782(2) |
| As1-Mn1-As1#5 | 92.318(6) | | |
| As1#4-Mn1-As1#5 | 180.0 | | |
| As1#1-Mn1-Mn1#6 | 56.392(4) | | |

___________________________________________________________________

Symmetry transformations used to generate equivalent atoms: #1 -x,-y,-z; #2 -x+1,-y+1,-z; #3 x-1,y-1,z; #4 -x,-y+1,-z; #5 x,y-1,z; #6 -x,-y,z-1/2; #7 -x,-y,z+1/2; #8 –x+1,-y+1,z+1/2; #9 x,y+1,z; #10 -x,-y+1,z+1/2; #11 x+1,y+1,z

**Table 4.** Anisotropic displacement parameters ($Å^2 \times 10^3$) for MnAs. The anisotropic displacement factor exponent takes the form: $-2\pi^2[\, h^2 a^{*2} U^{11} + ... + 2\, h\, k\, a^*\, b^*\, U^{12}\,]$

| Atom | $U^{11}$ | $U^{22}$ | $U^{33}$ | $U^{23}$ | $U^{13}$ | $U^{12}$ |
|---|---|---|---|---|---|---|
| Mn1 | 17(1) | 17(1) | 12(1) | 0 | 0 | 9(1) |
| As1 | 13(1) | 13(1) | 18(1) | 0 | 0 | 7(1) |

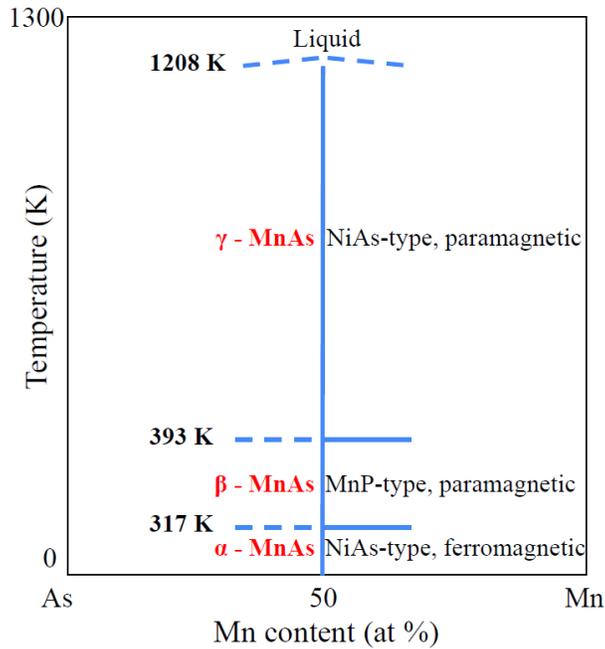

**Figure 1.** Schematic phase diagram of the As-Mn system in the vicinity of the stoichiometric 50:50 composition (adapted from Ref. [8]). Upon warming, bulk MnAs undergoes a first-order phase transition at ~317 K from a high-spin ferromagnetic to a low-spin paramagnetic state with a structural transition from the hexagonal α-phase (NiAs-type) to the orthorhombic β-phase (MnP-type). A second-order phase transition takes place at ~393 K, where the β-phase transforms to the γ-phase while remaining paramagnetic. The MnAs melts congruently (~1208 K) at ambient-pressure conditions.

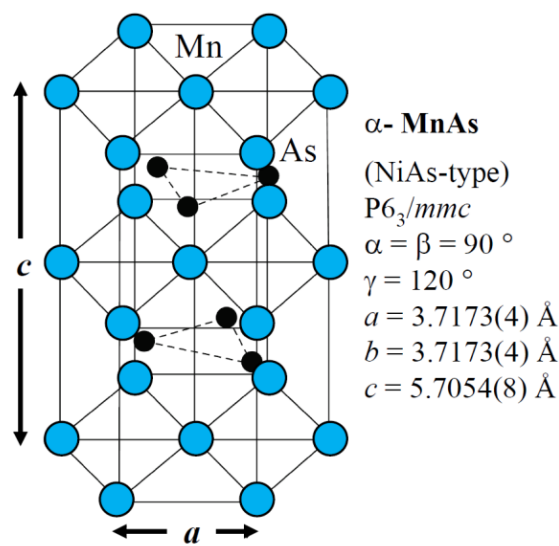

**Figure 2.** 3D view of the NiAs-type hexagonal structure ($P6_3/mmc$, $Z = 2$) of the α-MnAs phase. Mn and As atoms are shown as blue and black spheres, respectively.

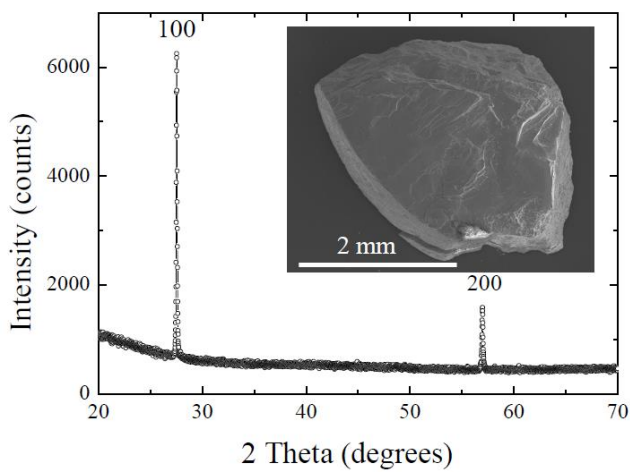

**Figure 3**. X-ray diffraction pattern showing the (*h00*) lines (main panel) and a scanning electron micrograph (inset) of a MnAs crystal grown under high-pressure conditions.

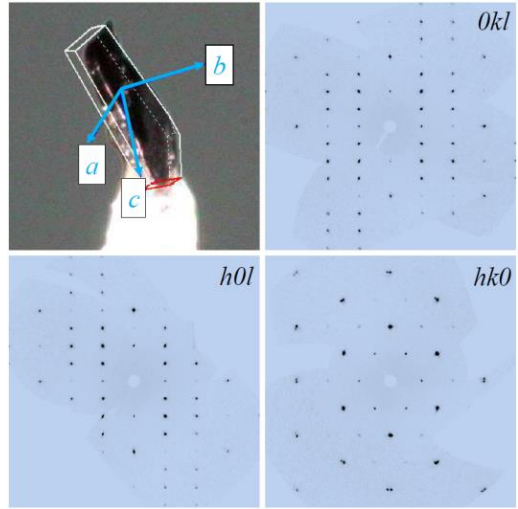

**Figure 4.** Image of a single MnAs crystal with unit cell axes and the reconstructed (*0kl*), (*h0l*), and (*hk0*) reciprocal space sections, as measured at room temperature. No additional phases, impurities, twins, or intergrowth crystals were detected.

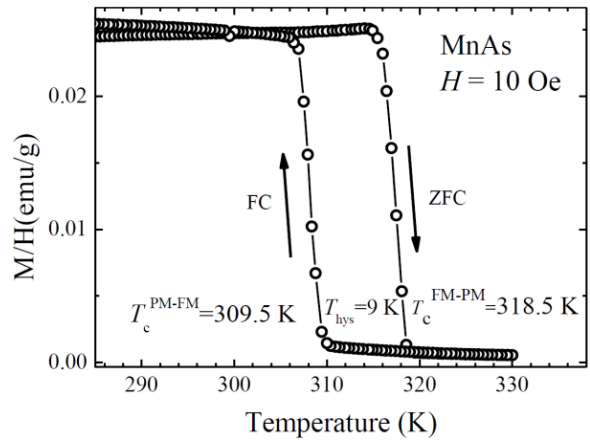

**Figure 5.** First-order phase transition and relevant hysteresis from a MnAs single crystal. Temperature dependence of the magnetization for a manganese arsenide single crystal in the magnetic field $H = 10$ Oe during heating (ZFC) and cooling (FC). ZFC and FC denote data measured in zero-field- and field-cooling, respectively.